\documentclass[aps,physrev,preprint,groupedaddress,citeautoscript]{revtex4-2}
\usepackage{xcolor}
\usepackage{amsmath,amssymb,amsfonts}
\usepackage[normalem]{ulem}
 \usepackage{physics}
 \usepackage{graphicx}
\usepackage{dcolumn}
\usepackage{bm}
\usepackage{hyperref}

\begin{document}


\title{Collective Excitonic Structure Governs Anomalously Weak Thermal Optical Dephasing in Conjugated Polymers}


\author{Henry~J.~Kantrow}
\affiliation{School of Chemical and Biomolecular Engineering, Georgia Institute of Technology, North Avenue, Atlanta, GA~30332, United~States}

\author{Elizabeth~Guti\'errez-Meza}
\affiliation{School of Chemistry and Biochemistry, Georgia Institute of Technology, 901 Atlantic Drive, Atlanta, GA~30332, United~States}

\author{Eric~R.~Bittner}
\affiliation{Department of Physics, University of Houston, Houston, Texas 77204, United~States}

\author{Hao~Li}
\affiliation{Institut Courtois \& D\'epartement de Physique, Universit\'e de Montr\'eal, 1375 Avenue Th\'er\`ese-Lavoie-Roux, Montr\'eal H2V~0B3, Qu\'ebec, Canada}

\author{Carlos~Silva-Acu\~{n}a}
\email[]{carlos.silva@umontreal.ca}
\homepage[]{https://silvascience.org}
\affiliation{Institut Courtois \& D\'epartement de Physique, Universit\'e de Montr\'eal, 1375 Avenue Th\'er\`ese-Lavoie-Roux, Montr\'eal H2V~0B3, Qu\'ebec, Canada}


\date{\today}

\begin{abstract}
Conjugated polymer aggregates exhibit optical decoherence in the presence of
strong vibronic coupling and substantial diversity in chemical structure,
solid-state organization, and excitonic character. Here, we use
coherence-detected and population-detected two-dimensional electronic
spectroscopies to examine the temperature dependence of the homogeneous optical
linewidth across a series of semiconducting polymers. Despite substantial
differences in molecular architecture and solid-state organization, all
polymers studied exhibit remarkably weak thermal linewidth scaling over the
measured temperature range. Comparison between complementary detection
modalities further shows that, while this weak temperature dependence is
robust, the absolute homogeneous linewidth depends on the measured observable.
This behavior reflects the different ways in which coherence- and
population-detected measurements project population relaxation and pure
dephasing onto the spectroscopic response. These results establish the weak thermal scaling of optical decoherence across a diverse series of conjugated polymers and show that its experimental manifestation must be interpreted in the context of the detection observable.
\end{abstract}


\maketitle

\section{Introduction}

A central challenge in materials science is to understand how molecular
organization governs electronic function across length scales. Conjugated
polymers provide a particularly rich setting in which to address this problem:
delocalized $\pi$ electrons mediate their optical and electronic
response~\cite{Pope1999}, while their solid-state organization is shaped by
chemical structure, regioregularity, molecular weight, and processing
history~\cite{Noriega2013,koch2013impact,Wunderlich:1976aa}. Polymer chains
therefore assemble into microstructures with widely varying degrees of order
and inter- and intrachain organization, creating multiple pathways for
electronic coupling between backbone
chromophores~\cite{Spano2010AccChemRes,Spano:2014aa,Spano2018ChemRev}.

These interactions generate photophysical aggregates: collective electronic
states formed through coherent coupling of transition dipoles on neighboring
chromophores, as originally described by Kasha~\cite{kasha1950}. Their optical
properties consequently reflect not individual molecules, but interacting
ensembles whose electronic structure depends on molecular organization. This
connection is well established in the spectral signatures of conjugated
polymers, where interchromophore coupling, vibronic interactions, and energetic
disorder collectively determine absorption and
emission~\cite{Spano2005JChemPhys,Spano2007,Silva2007PhysRevLett,Clark:2009aa,spano2009JChemPhys,Paquin:2013aa,yamagata2012interplay}.
A less understood question is whether molecular organization also determines
how rapidly an optically excited state loses its initially well-defined
relationship to the driving light field as it interacts with its surroundings.

This loss of optical coherence is described by the dephasing time $T_2$ and
sets the homogeneous linewidth of an optical transition and provides a dynamical measure of how an
excitation interacts with molecular vibrations, structural fluctuations, and
other degrees of freedom in its
environment~\cite{Mukamel1995,Tokmakoff2000,Pascal2017PRB}. The temperature
dependence of the homogeneous linewidth is particularly informative because
thermal excitation changes these environmental fluctuations and therefore
reveals which interactions contribute to the loss of optical
coherence~\cite{osad1991optical}. Conjugated polymers provide a stringent
setting in which to examine this relationship because they combine strong
vibronic activity with complex intra- and interchain electronic interactions
and substantial structural disorder.

In conjugated polymers, these interactions coexist to produce mixed HJ
excitonic character rather than the limiting behavior of a purely H- or
J-type molecular aggregate~\cite{yamagata2012interplay,Yamagata2014JPhysChemC,Spano:2014aa,Spano2018ChemRev,balooch2020vibronic,chang2021hj}.
Resonance-Coulomb and charge-transfer interactions, conformational disorder,
and intermolecular organization can modify the degree of H/J admixture,
exciton delocalization, and coupling to the molecular environment. The
diversity of polymer microstructures therefore provides an opportunity to ask
whether temperature-dependent optical dephasing exhibits common behavior
despite substantial variation in chemical structure, mesoscale organization,
and excitonic character. Although nonlinear optical spectroscopy has been
applied extensively to conjugated polymers, the temperature dependence of
their homogeneous linewidth has not been established systematically across
such a broad polymer series.

Here, we address this question using coherence-detected and
population-detected two-dimensional electronic spectroscopies to determine
temperature-dependent homogeneous linewidths across five conjugated
polymers---P3HT, P3HHT, PBTTT, PCE11, and N2200. These materials span
substantially different forms of molecular and mesoscale organization. We find
weak thermal scaling of the homogeneous linewidth across the entire series.
Where both detection modalities are available, the absolute linewidths differ,
while their weak temperature dependence persists, demonstrating that the
magnitude of the measured homogeneous linewidth depends on the detected
observable~\cite{paiva2026detection}.

\section{Results}

\subsection{Temperature-dependent homogeneous linewidth}

\begin{figure}[tb]
    \centering
    \includegraphics[width=1\columnwidth]{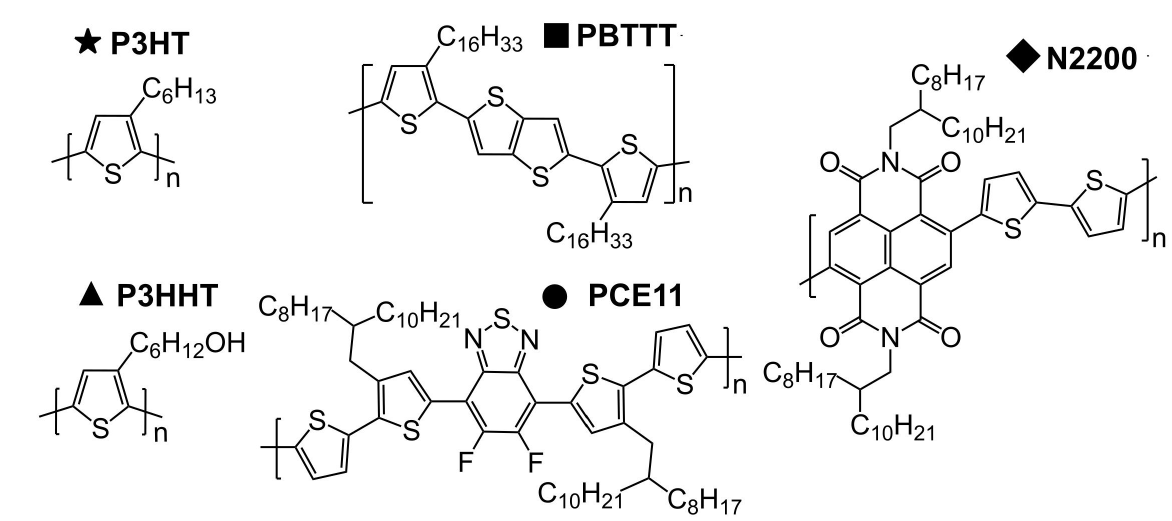}
   \caption{\textbf{Conjugated-polymer series studied in this work.}
Molecular structures of P3HT, P3HHT, PBTTT, PCE11, and N2200.
Symbols identify the corresponding materials in
Fig.~\ref{fig:linewidth-vs-temp}.}
    \label{fig:materials}
\end{figure}

The conjugated polymers studied here are shown in
Fig.~\ref{fig:materials}. P3HT, P3HHT, PBTTT, N2200, and PCE11 span a broad
range of backbone structures, conformational flexibility, donor--acceptor
character, and solid-state organization. P3HT is a flexible-chain polymer~\cite{heffner1991molecular} that
provides a canonical semicrystalline reference, forming a two-phase
microstructure of crystalline and amorphous regions at the molecular weight
used here (125\,kg\,mol$^{-1}$)~\cite{koch2013impact}. P3HHT retains the same
thiophene backbone while incorporating hydroxylated side chains that promote
hydrogen bonding and modify backbone organization~\cite{nicolini2021low}.

PBTTT represents a distinct structural regime, forming liquid-crystalline-like
assemblies associated with the thienothiophene-containing backbone and
interdigitation of its long alkyl side chains~\cite{McCulloch2006,kline2007critical,delongchamp2008molecular,ratcliff2025terra}.
N2200 and PCE11 extend the comparison to donor--acceptor copolymers, in which
intrachain charge-transfer character modifies the balance of H- and J-like
excitonic interactions~\cite{balooch2020vibronic,chang2021hj}. N2200 combines
torsional flexibility associated with its thiophene units with strong
donor--acceptor character arising from the electron-withdrawing naphthalene
diimide unit~\cite{chen2009naphthalene,giussani2013molecular}. PCE11 similarly exhibits donor--acceptor character associated with its
benzothiadiazole unit, together with noncovalent heteroatom interactions that
influence backbone planarity and solid-state
packing~\cite{wang2018bulk,deng2023determinant}.
The series therefore samples substantially different forms of molecular and
mesoscale organization within the broader class of conjugated-polymer
aggregates. We use this structurally diverse series to examine whether these differences
are accompanied by corresponding differences in optical coherence and its
temperature dependence.

A central challenge in determining the homogeneous linewidth of molecular
aggregates is that the optical coherence time $T_2$ contains contributions
from both population relaxation and phase-randomizing environmental
fluctuations. A useful starting point is the standard decomposition
\begin{equation}
\frac{1}{T_2}
=
\frac{1}{2T_1}
+
\Gamma_\phi,
\label{eq:T2decomp_main}
\end{equation}
where $T_1$ is the population-relaxation time and
$\Gamma_\phi=1/T_2^{*}$ is the pure-dephasing rate. Population relaxation
describes processes that remove amplitude from the initially prepared optical
state, whereas pure dephasing describes loss of phase memory through
fluctuations of the transition energy without a corresponding change in
excited-state population.

We determine the homogeneous linewidth using two complementary forms of
two-dimensional electronic spectroscopy. The first is two-dimensional
coherent spectroscopy in a non-collinear, phase-matched geometry implemented
using the coherent optical laser beam recombination technique (COLBERT),
originally developed by Turner~\textit{et~al.}~\cite{Turner:2011aa} and used
extensively in our previous work on conjugated-polymer
photophysics~\cite{zheng2024unveiling}. Because the rephasing spectrum
separates homogeneous and inhomogeneous contributions to the optical
lineshape, COLBERT provides direct experimental access to the decay of the
optically prepared coherence.

\begin{figure}[tb]
    \centering
    \includegraphics[width=0.8\columnwidth]{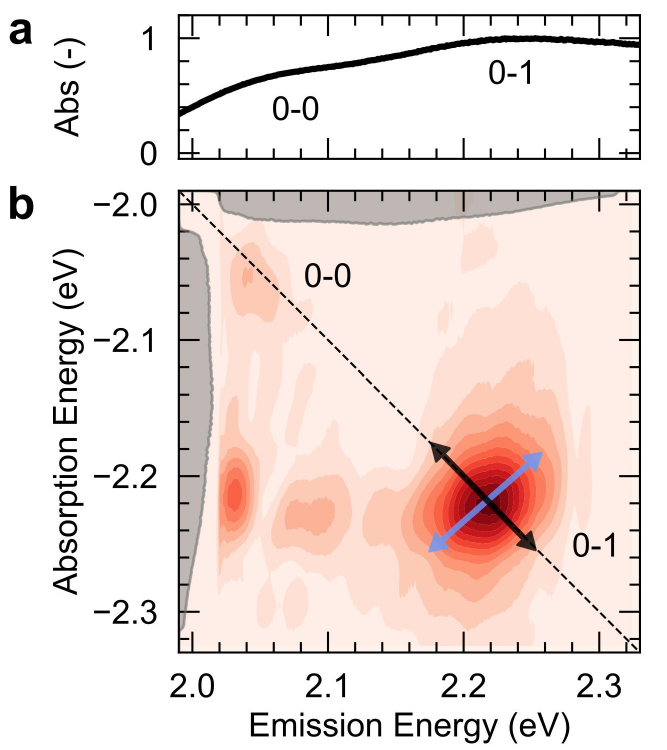}
    \caption{\textbf{Extraction of the homogeneous linewidth from
    coherence-detected two-dimensional spectroscopy.}
    (a) Linear absorption spectrum of PBTTT, with the 0--0 and 0--1 vibronic
    transitions near 2.08\,eV and 2.22\,eV indicated. (b) Representative
    one-quantum rephasing spectrum of PBTTT at a population time of 20\,fs.
    Fits to diagonal (black arrow) and antidiagonal (blue arrow) cuts are used
    to separate inhomogeneous and homogeneous broadening. The excitation bandwidth is indicated in gray.}
    \label{fig:COLBERT_2DPL_comparison}
\end{figure}

The COLBERT spectrum of PBTTT provides an illustrative example of the
linewidth extraction. The excitation bandwidth covers the 0--0 and 0--1
vibronic transitions observed in the linear absorption spectrum
(Fig.~\ref{fig:COLBERT_2DPL_comparison}a). The corresponding rephasing
spectrum displays the principal diagonal features together with cross-peaks
between the vibronic transitions
(Fig.~\ref{fig:COLBERT_2DPL_comparison}b). Homogeneous and inhomogeneous
broadening are separated through antidiagonal and diagonal cuts of the
rephasing spectrum, respectively, following the lineshape analysis described
by Siemens~\textit{et~al.}~\cite{siemens2010resonance}.

The second approach is photoluminescence-detected two-dimensional electronic
spectroscopy (2DPL), introduced by Tekavec~\textit{et~al.}~\cite{Tekavec2007}
and implemented in our previous studies of conjugated
polymers~\cite{Pascal2017PRB,li2016probing,gutierrez2021frenkel}. 
In 2DPL, the nonlinear excitation sequence is encoded in an incoherent
population observable rather than in a phase-matched emitted field.
Homogeneous linewidths can likewise be extracted from the diagonal and
antidiagonal structure of the rephasing lineshape.

The distinction between the two detection modalities is important. Although
both originate from the same optically driven excitonic dynamics, a
population-detected signal encodes the nonlinear response through an
excited-state population that evolves before photoluminescence detection. It
can therefore weight relaxation pathways differently from a
coherence-detected measurement. Population-detected nonlinear spectra can
additionally contain incoherent population-mixing
contributions~\cite{gregoire2017incoherent}; these can be identified and
separated from the nonlinear coherent response using procedures described
previously~\cite{bargigia2022identifying}. More generally, the homogeneous
linewidth obtained experimentally should be understood as the linewidth
associated with the particular detected observable rather than as a
measurement-independent material constant~\cite{paiva2026detection}.

\begin{figure*}[t!]
    \centering
    \includegraphics[width=0.90\linewidth]{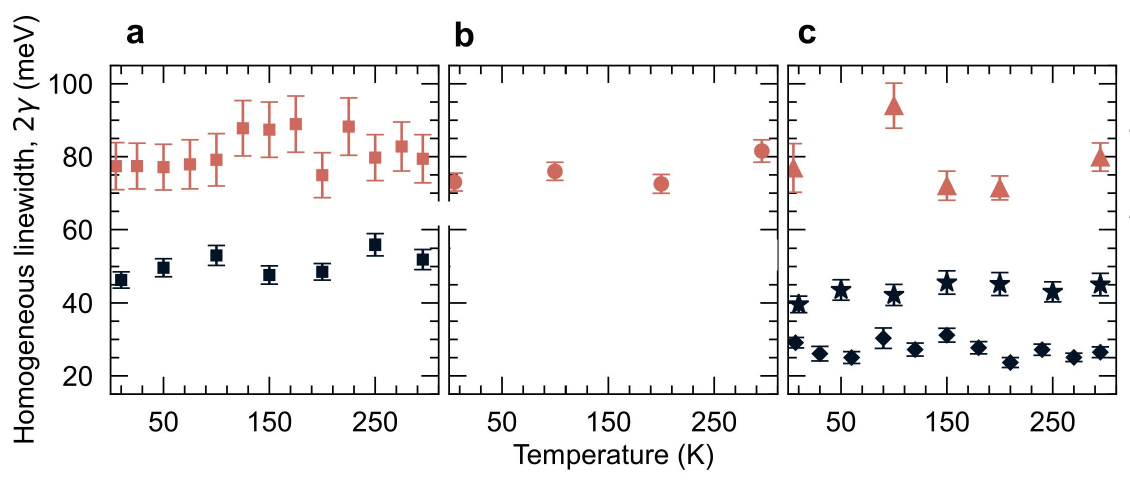}
   \caption{\textbf{Temperature-dependent homogeneous linewidths of the
conjugated-polymer series.}
Homogeneous linewidth $2\gamma=\hbar T_2^{-1}$ as a function of
temperature for the polymers shown in Fig.~\ref{fig:materials}.
(a) PBTTT measured by coherence-detected COLBERT and population-detected
2DPL. (b) PCE11 measured by 2DPL. (c) P3HHT, P3HT, and N2200 measured by
COLBERT and/or 2DPL as indicated. Symbols correspond to those defined in
Fig.~\ref{fig:materials}.}
    \label{fig:linewidth-vs-temp}
\end{figure*}

The homogeneous linewidths,
\begin{equation}
2\gamma=\hbar T_2^{-1},
\end{equation}
obtained for the polymers in Fig.~\ref{fig:materials} are shown as a function
of temperature in Fig.~\ref{fig:linewidth-vs-temp}. We first compare COLBERT
and 2DPL measurements of PBTTT
(Fig.~\ref{fig:linewidth-vs-temp}a). The homogeneous linewidth obtained from
2DPL is systematically larger than that obtained from COLBERT. Despite this
offset, both measurements exhibit similarly weak thermal scaling over the
experimentally accessible range. PBTTT therefore illustrates that the
absolute linewidth can depend on the detection observable even when its
temperature dependence is preserved.
Importantly, the difference in absolute linewidth is not accompanied by a
correspondingly strong difference in its temperature dependence. Both
measurements of PBTTT exhibit only weak thermal scaling over the experimentally
accessible range. The comparison therefore separates two aspects of the
measurement: the absolute homogeneous linewidth depends on how the nonlinear
response is detected, whereas its weak temperature dependence persists across
the two modalities.

The same weak temperature dependence extends across the remainder of the
polymer series. PCE11 exhibits little variation of its 2DPL linewidth over the
measured range (Fig.~\ref{fig:linewidth-vs-temp}b), while P3HHT, P3HT, and
N2200 similarly show weak thermal scaling in the available COLBERT and 2DPL
measurements (Fig.~\ref{fig:linewidth-vs-temp}c). The absolute linewidths span
a substantially broader range than their temperature dependence: values across
the series extend from approximately 20\,meV to approximately 90\,meV, whereas
the linewidth of each material changes only weakly over its measured
temperature interval.

N2200 exhibits the narrowest homogeneous linewidth in the series, reaching
approximately 20\,meV in the COLBERT measurement. The origin of the substantial
differences in absolute linewidth among the polymers cannot be established from
the present measurements alone and would require a detailed treatment of their
aggregate excitonic structures. Here, the relevant observation is that these
large differences in linewidth magnitude are accompanied by consistently weak
thermal scaling.
Establishing the microscopic origin of the differences in absolute linewidth among the polymers requires a detailed treatment of their aggregate excitonic structure and lies beyond the scope of the present analysis. The important result here is that the substantial variation in absolute linewidth across chemically and structurally distinct polymers is accompanied by consistently weak thermal
scaling.


\section{Discussion}

The most striking feature of the measurements is the consistently weak
temperature dependence of the homogeneous linewidth across the conjugated
polymer series. P3HT, P3HHT, PBTTT, PCE11, and N2200 differ substantially in
chemical structure, conformational flexibility, donor--acceptor character, and
solid-state organization, yet none exhibits strong thermal broadening over the
measured temperature range. At the same time, their absolute homogeneous
linewidths span approximately 20--90\,meV. The magnitude and temperature
dependence of the linewidth therefore constitute distinct experimental
characteristics: polymers with substantially different optical coherence
times can nevertheless exhibit similarly weak thermal scaling.

Comparison between detection modalities provides a second distinction. Where
both coherence-detected COLBERT and population-detected 2DPL measurements are
available, the absolute linewidths need not coincide even though their
temperature dependences are similar. This behavior is consistent with the
view that the measured homogeneous linewidth reflects a projection of
exciton--environment dynamics onto the detected spectroscopic
observable~\cite{paiva2026detection}. Coherence detection and population
detection interrogate the nonlinear response through different pathways and
can therefore weight the dynamical processes contributing to the measured
lineshape differently.

The persistence of weak thermal scaling across both the polymer series and
the available detection modalities places an important constraint on a
microscopic description of optical decoherence in these materials. Differences
in molecular structure, mesoscale organization, and absolute linewidth do not
produce comparably strong differences in thermal broadening. Determining how
specific features of aggregate excitonic structure give rise to this behavior
requires a detailed treatment of exciton--vibrational coupling and relaxation
pathways beyond the scope of the present work.

\begin{acknowledgments}
HL and CSA acknowledge funding from the Government of Canada (Canada Excellence Research Chair CERC-2022-00055). CSA acknowledges support from the Institut Courtois, Facult\'e des arts et des sciences, Universit\'e de Montr\'eal (Chaire de recherche de direction de l'Institut Courtois) and from the Natural Science and Engineering Research Council of Canada (NSERC Discovery Grant RGPIN-2024-05893). ERB acknowledges funding from the National Science Foundation (CHE-2404788), Robert A.\ Welch Foundation (E-1337), the Department of Energy supported this research through Award No. 11937-PO147716. ERB gratefully acknowledges funding from IVADO for a Visiting Professorship at the Institut Courtois, Universit\'e de Montr\'eal. 
\end{acknowledgments}

\section*{Author declarations}

\subsection*{Conflict of Interest}
The authors have no conflicts to disclose.

\subsection*{Use of Generative Artificial Intelligence}
In compliance with institutional guidelines of the Universit\'e de Montr\'eal, generative artificial intelligence tools were used to assist with the editing of language and stylistic refinement of parts of the manuscript and to assist in the synthesis of the literature. These tools were not used to generate scientific content, perform analysis, or influence the interpretation of results. All content has been reviewed and validated by the authors, who assume full responsibility for the manuscript.

\section*{Data availability}
In accordance to the Data Management Plan of the Canada Excellence Research Chair in Light-Matter interactions (\url{http://hdl.handle.net/1866/33427}), the numerical data and code that support the findings of this study are openly available in the Borealis Dataverse Repository at [URL to be included prior to publication].

\end{document}